\documentclass{emulateapj}

\slugcomment{ApJ accepted June 6, 2006}

\shorttitle{IR constraint to a Companion of Vega}
\shortauthors{Hinz et al.}

\begin{document}

\title{Thermal Infrared Constraint to a Planetary Companion of Vega  \\ 
         with the MMT Adaptive Optics System 
         \footnotemark[1]}
       
\footnotetext[1]{Observations reported here were obtained at the MMT Observatory, a joint facility of the University of Arizona and the Smithsonian Institution.}  

\author {Philip M. Hinz, A. N. Heinze, Suresh Sivanandam, Douglas L. Miller, Matthew A. Kenworthy,  			Guido Brusa, Melanie Freed, and J.R.P. Angel}
\affil{ Steward Observatory, University of Arizona, 933 N. Cherry Ave., Tucson, AZ 85721}

\begin{abstract}
Vega may have a massive companion in a wide orbit, as evidenced by structure in its cold dust debris.  We have tested this hypothesis by direct imaging with adaptive optics in the M band.  The observations were made with a newly commissioned thermal infrared camera, Clio, on the 6.5 MMT AO system with low-background deformable secondary.  The observations constrain a planet to be less than 7 M$_J$ at the approximate position angle expected from the dust structure and at a radius $>$ 20AU (2.5 arcsec) .  This result is more stringent than similar previous near-infrared observations of Vega, that achieve limits of 20 and 10 M$_J$ at separations of 7 arcsec.  The higher sensitivity is due both to the more favorable contrast of gas giant planets at M band and to the higher Strehl and more stable point spread function at longer wavelengths.  Future L' or M band observations could provide a powerful approach for wide separation planet detection, especially for cooler, and thus older or less massive planets. The natural best targets are nearby stars where planets in the range of 5-15 M$_J$ and as old as several Gyr are expected to be detectable with this technique.

\end{abstract}


\section{Introduction}

Direct detection of extrasolar planets is a highly desirable goal for a range of reasons.  Although planet detection through its gravitational influence on its star has yielded important information about planets around other stars, many parameters of a planet are most easily derived from its spectral energy distribution (SED) including its temperature, size, and composition.  For planets in orbit beyond approximately 5-10 AU, the length of time needed to detect a planet through indirect means also becomes prohibitive.  Yet these are precisely the radii where we see massive planets in our own solar system.

Initial success in the area of direct detection is just now occurring, although the detectable objects are still significantly different from what we think of as a typical solar system.  Measurements of secondary eclipses of transiting hot Jupiters can begin to constrain the temperature and albedo of these objects (Deming et al. 2005,  Charbonneau et al. 2005, Rowe et al. 2005).  For younger stars and wider separations faint companions have been detected (Chauvin et al. 2004, Neuhauser et al. 2005) which are consistent with being planetary mass objects. However, these objects are unlikely to have been formed by either core accretion or gravitational instability (Boss 2006) suggesting that we are either seeing scattered planets or low mass objects that have formed by fragmentation.

Direct detection searches are typically focusing on very young stars where giant planets are early in the process of cooling and contracting and thus are expected to be relatively bright in the near infrared ( see Metchev et al. 2004, Oppenheimer et al. 2004, Mugrauer et al. 2005, Lowrance et al. 2005, and Biller et al. 2005 as examples of ongoing surveys).  Of these bands the H band is particularly attractive due to the methane absorption feature which creates an identifiable spectral feature, as well as its relative brightness compared to the K band.  The majority of the direct detection plans with large telescopes focus in this spectral region, utilizing techniques such as simultaneous spectral difference imaging (Marois et al. 2000, Biller et al. 2005) and higher order adaptive optics (Oppenheimer et al. 2004, Macintosh et al. 2003a) to optimize the achievable contrast.  Several observatories (see Gratton et al. 2004 as an example) are currently designing ambitious near-infrared instruments to pursue the detection of young planets.

Although near-infrared sensitivity from the ground is better than at longer wavelengths, a significant contrast advantage can be achieved by looking for cool giant planets closer in wavelength to the peak of their expected SED.  Giant planets have very non-blackbody SEDs, requiring a knowledge of their spectral characteristics to best choose an optimum wavelength.  Additionally, the sensitivity of a ground-based system is dominated by the spectral dependence of the atmospheric transparency and brightness, limiting possible detection to discrete spectral windows.

It has long been known  that Jupiter's SED displays an anomolous peak at approximately 4-5 microns (Gillett, Low, and Stein  1969). This is due to the lack of absorption features in this spectral region which allows the observation of thermal emission from much deeper in the planet's atmosphere than at other wavelengths.  A similar, broader peak can be seen in T type brown dwarfs (Oppenheimer et al. 1998) and is an expected generic feature of objects for objects with effective temperatures below approximately 1200 K, as modeled by Burrows et al. (1997, 2003) and Baraffe et al. (2003).  The high flux appears in the L' and M bands for hotter object and narrows to a feature which is well matched to the M band window for the coolest objects.  The relative flux in L' versus M bands for hotter objects, as measured by Golimowski et al. (2004) appear to be dependent on the amount of carbon monoxide in the atmosphere due to non-equlibrium effects, lessening the expected brightness at M compared to equilibrium models. Such effects may determine whether the M or L' atmospheric window is preferable, but the existence of a broad hump in the SED in the 4-5 micron region appears to be secure both from model predictions and observed cool objects.

To make use of this expected contrast enhancement it is necessary to be able to detect the relatively faint planet flux in the presence of background from the sky as well as any warm optics in the system.  The relatively small expected separations require diffraction-limited performance of the optics, which in turn requires an adaptive optics system to correct for atmospheric turbulence.  Typical AO systems add 5-10 warm surfaces. For wavelengths longward of approximately 2 microns the infrared glow from the warm optics can dominate the background and, thus, the noise of an optical system.  For planet detection at longer wavelengths an optimum system is one which is integrated into the telescope itself, such as the deformable secondary mirrror of the MMT AO system (Riccardi et al. 2002, Brusa et al. 2003).

Optimal targets for direct detection of extrasolar planets are stars which are both young and nearby.  Unfortunately most radial velocity targets do not satisfy both of these criteria.  Target stars are typically chosen to avoid potential noise in a Doppler signal from stellar activity (Wright et al. 2004).  For young stars, a possible alternative indicator of the existence of a planetary companion is substructure in a debris disk (Roques et al. 1994, Liou \& Zook 1999). Giant planets will tend to set up resonances in debris material around a star with the chracterisitics of the resonances dependent on parameters of the planet's mass and obrbit  (Kuchner \& Holman 2003).  If this substructure can be modeled properly it may be possible to use such information to guide where and which stars to search for planetary companions.

In this letter we present initial observations with Clio (Freed et al. 2004), a camera designed specifically to detect giant planets through their 3-5 micron infrared radiation.  We have used the imager to constrain the existence of a planetary companion to Vega, at the orientation expected from resonances in the cold debris disk.    Section 2 describes the instrumental setup.   In section 3 we present the observations.  Section 4 details the data reduction and achievable limits versus separation.  In Section 5 we discuss the implications of these initial observations for future planet detection.  

\subsection{The Vega System}

Vega provides an interesting target for planet searches as one of the nearest young stars with a debris disk indicative of planetesimals. For the purpose of planet models used in this paper we adopt an age of 300 Myr and a distance of 7.8 pc consistent with evolutionary tracks (Song et al. 2001) and Hipparcos astrometry. The debris disk around Vega has been intensively studied since its discovery with IRAS (Aumann et al. 1984).   It appears that we are viewing a cold (Backman \& Paresce 1994), probably transiently bright (Su et al. 2005) dust disk equivalent to the Kuiper belt at nearly pole-on orientation around Vega.   Sub-mm and millimeter observations of Vega have revealed a double-lobed enhancement to the ring (Holland et al., 1998, Koerner et et al., 2001, Wilner et al., 2002, hereafter W02).    The offset nature of the lobes seen at long wavelengths have been modeled by W02. They reproduce the observed structure for a planet which is approximately 90 $\deg$ from the line connecting the lobes, on the same side of the star as the offset of the lobes.  They suggest that the planet is 7" from the star in the northwest direction.

Interestingly, the observed structure seen around Vega is similar to one of the four broad classes of dust structure described by Kuchner and Holman (2003) which can arise from giant planet perturbations to debris disks.  The class which fits the double-lobed pattern of Vega is indicative of a high-mass planet on an eccentric orbit. The mass of the planet is not well-constrained by the observable resonant structure, although W02 achieve good correspondence between their model and the observations using a 3 M$_J$ planet.  Another hypothesis for the lobed structure has been proposed by Wyatt (2003). where the resonance arises from outward planetary migration of a less massive (Neptune-mass) planet from 40 to 65 AU. Direct imaging has the potential to detect massive planets around Vega, potentially providing an important constraint to modelling of the resonant structure.

Detection of a companion to Vega has previously been attempted with the Palomar system in H band. Metchev et al. (2003) set a 5 sigma limit of approximately 20 M$_J$ based on their observations at the separation of the expected perturber.  Macintosh et al. (2003) also used Keck at K band to carry out a similar observation. They estimate their 5 sigma limit corresponded to 10 M$_J$ at the expected separation of the planet. More recent deep integrations have been achieved by Marois et al. (2006) reaching a limit of approximately 4 M$_J$ at a separation of 7 $\arcsec$.

\section{Instrumental Setup}

The observations described below made use of two recent developments for the MMT telescope:  a deformable secondary adaptive optics system and a high well depth, high duty cycle Indium Antimonide based camera system.  We describe the advantages of these approaches below.

\subsection{The MMT deformable secondary AO system}

The MMT Adaptive Optics (AO) system makes use of the world's first deformable secondary mirror (Riccardi et al. 2002, Brusa et al. 2003), allowing integration of the AO into the telescope itself (Wildi et al. 2003).  The deformable secondary is undersized to form the stop of the system, and provide an effective aperture of 6.35 m diameter. Secondary undersizing is a proven technique to provide efficient baffling of warm background radiation from the telescope structure.  Infrared light encounters only two warm reflecting surfaces, the primary and secondary mirror before entering the cryogenically cooled camera which eliminates background contributions from these optics.  The result is an infrared optimized system which is also capable of diffraction-limited imaging  (a full width at half maximum of 0.13 arcseconds at L' and 0.16 arcseconds at M band).  Compared to conventional adaptive optics systems this results in improved sensitivity.  Lloyd-Hart (2000) shows that for typical parameters expected at the MMT, observations with an optimized system can achieve the same signal in one third to one half of the integration time that would otherwise be needed using a conventional adaptive optics system at L' and M bands. 

 Atmospheric turbulence is sensed using visible light reflected from the entrance window of the cryostat.  The telescope pupil is imaged onto a CCD-based Shack Hartmann wavefront sensor operating at 550 Hz.  The resulting slopes are reconstructed to measure and correct the first 56 Zernike modes of the turbulent wavefront.  The system typically delivers Strehls of approximately 25\% at H band in median seeing (Miller et al. 2004).  Scaled to M band this is a Strehl of 85\%. 

\subsection{Clio: a 3-5 micron imager}

Clio (Freed et al. 2004, Sivanandam et al. 2006) is an imager designed for obtaining high spatial resolution images with optimum efficiency at L' and M band.  The favorable contrast for planets at M band is at least partially offset by the high background in this atmospheric window.  Our initial estimates of the relative sensitivity in these two bands suggests the L' band magnitude limit is approximately 2 magnitudes better than the M band limit, consistent with the higher background at M band compared to L'. This is a good match to the expected colors for giant planets.  For example, models of giant planets (Baraffe et al. 2003, Burrows et al. 2003) predict that a planet will be roughly 2 magnitudes brighter in M band versus L' for an effective temperature around 350 K.  The coincidence of these two factors, a fainter L' limiting magnitude and a red L'-M color for a typical planet,  provides favorable conditions for confirmation of any substellar companion.  While common proper motion will still be important to verify physical association, confirmation via color is a strong secondary indicator that an object is indeed a planetary mass object.  

High efficiency observations with Clio are aided by an Indium Antimonide (InSb) detector, from Indigo Systems, Inc.  which has a high well depth compared to more typical astronomical InSb detectors.  This feature allows efficient use in the M band where high backgrounds can typically swamp detectors with lower well depths.   The detector has a measured well depth of 3 million photo-electrons (Sivanandam et al. 2006).  For the initial observations the frame rate was only 4 Hz.  This resulted in a duty cycle of roughly 50\% for the 120 ms frame time used in the observations of Vega.  The camera readout has been since improved to 20 Hz.  This will allow a higher observational efficiency in the future with Clio.

The camera optics reimage the telescope to a cold mask which eliminates the surrounding radiation from warm telescope structure.  The f/15 Cassegrain focus is reimaged at f/20 onto the 320x256 pixels, each 30 microns in width.  The resulting plate scale is approximately 0.05 arseconds per pixel.  The detector is oriented so that the long direction is in the elevation direction when the rotator offset is zero. 

\section{Observations}

Observations of Vega were taken with Clio and the MMT AO system on June 21, 2005, during initial tests of the Clio camera.  The filter used was from OCLI (now JDS Uniphase) with half power points at 4.35 and 4.95 microns, slightly shorter than a standard M band filter.  L' band images were also taken, but were later found to be contaminated by a filter misalignment in the camera and are not presented here.  Table 1 lists the data taken. Individual images, each 120 ms integration time, were coadded in computer memory.  One hundred images were combined for each frame, resulting in an integration time of 12 s for each frame. Fifty-six frames were acquired in closed loop with sufficiently high Strehl to use in the final image, resulting in a total integration time of 672 s. The telescope was periodically nodded by 5 arcseconds in the direction to move Vega vertically on the array, to allow for background subtraction. The star was placed on the lower left or upper left quadrant of the array for the two positions.  This allowed roughly 220 pixels or 11 arseconds of field on the west side of the star, where the hypothetical perturbing companion has been proposed to be.   The instrument derotator was turned off which resulted in slow sky rotation of the field.  Since we observed Vega well after transit this amounted to no more than 10 degrees rotation during the course of all the observations.  For roughly half the frames the derotator was turned from an offset of zero to an offset of -30 degrees.  This was done to check the approach of PSF subtraction via rotator offset.

\section{Data Reduction and Analysis}

Data analysis was carried out with custom C software developed for reduction of Clio data.  The software is being developed to provide relatively automated reduction of Clio data optimized for faint companion detection.  Initial background removal was achieved via subtraction of each frame with its nod pair. Although the images were saturated inward of approximately 0.3 arcseconds, boxcar smoothing the resulting images with a 25$\times$25 pixel aperture allowed a valid centroid to be calculated.  The calculated centroid for the positive and negative nod image in each frame was used to shift each image.  The parallactic angle was calculated from the time taken for each frame and used to derotate the images before final combination. Additional processing to account for drifts in detector bias, and noise in individual columns were carried out and are described further by Heinze et al. (2006). The resulting image has a total integration time on Vega of 672 seconds.  A broad halo and diffraction pattern from Vega dominates the resulting image.  To remove this the image was boxcar smoothed using a Gaussian kernel with FWHM of 5 pixels, and the resulting image was subtracted from the original to create an unsharp masked image.  Figure 1 shows the resulting image. The most sriking feature of the data is the number of Airy rings detectable in the data, a characteristic of the high Strehl achieved at M band with adaptive optics.

Noise contours were generated from the image in order to quantify where we are sensitive to what level of companions and are shown in Figure 2.  The boundaries between each successive greyscale are, from faintest to brightest, 1.13, 1.72, 2.54, 16, and 100 mJy.  For the models of Burrows et al. (2003) this corresponds to a limit of 7, 10, 14, and 26 M$_J$ for the four lowest contours respectively.  For regions of overlap of all of the frames, the approximate 5 sigma limit to a planet is 0.78 mJy, which corresponds to a mass limit of 6 M$_J$.  The data show that the noise is dominated by sky background outside of approximately 2.5 arcseconds. Figure 1 also shows the insertion of synthetic planets into the data.  Inspection of these planet images reveals that the actual limiting magnitude might be slightly higher than the 0.8 mJy formal calculation, possibly due to some amount of correlation in the detector pixels, an artifact of column noise which is only partially removable. However, the 1.1 mJy sources appear detectable, allowing us to feel confident that we would detect a planet down to approximately 7 M$_J$ in the darkest contour shown in Figure 2.

\subsection {Optimal PSF subtraction}

To test various ways of removing the diffracted or aberrated light from the star we investigated unsharp masking, Point SPread Function (PSF) subtraction via rotator offset changes, and psf subtraction via sky rotation. This data reduction was carried out independently of the data reduction described above, allowing an independent check on the sensitivity limit.  The scripts were developed using the Perl Data Language \footnote[1]{see http://pdl.perl.org}.  Separate coadded frames were generated using frames 25-48, and 51-70 respectively for images at 0 and -30 degrees of the telescope rotator offset.  No rotation of the images was carried out for this data reduction, since the main goal was to determine the stability of the PSF not use the images for faint companion detection.  Figure 3 shows the two resulting images at each rotator offset.  The images were unsharp masked using a 0.25 $\arcsec$  boxcar. The static speckle pattern in the image is clearly tied to the rotator offset, suggesting the source is the telescope optics, rather than the camera optics.  This results in poor subtraction if the rotator offset is used and suggests sky rotation will be the preferred way to obtain multiple images for later PSF subtraction.  This corroborates the results of Marois et al. (2006) that Angular Differential Imaging (ADI) is a useful approach to obtain good PSF subtraction with an adaptive optics system.

 Although our images were not taken during significant sky rotation, it is possible to simulate this process to evaluate the achievable limits. To do so, we rotated the second image in Figure 3 by 30 degrees so that the static pattern of the two images are aligned. These images were taken roughly 40 minutes apart and across a change in elevation of the telescope of 7 degrees, similar parameters to what would be expected when taking multiple images during a rapid change in parallactic angle of the object.  Good subtraction of the PSF is obtained using this approach.  

To quantify the achievable contrast limit in an image, a boxcar smoothing of the image over a 3x3 pixel box was applied and annuli around the central pixel were analyzed to estimate the variation in flux in an aperture approximately the size of the PSF.  For each annulus the standard deviation of all the included pixels was calculated.  A threshold of five times the standard deviation is a reasonable estimate of the limiting contrast versus separation. A plot of these values is shown in Figure 4 for three separate images. The solid line shows the limit for a raw PSF.  The dashed line shows the limit for an unsharp masked (5x5 pixel boxcar smoothed) image.  The dot-dashed line shows the result for a PSF-subtracted image which has also been unsharp masked. 

 Figure 4 verifies that PSF subtraction using sky rotation will be a useful approach for high dynamic range imaging in M band.  We expect to achieve contrast limits of roughly 11 magnitudes at 1 arcsecond and 13 magnitudes by 1.5 arcseconds.  While sky rotation did not allow us to take sufficient data to use this method for Vega, these initial data suggest that observing an object through significant sky rotation will be the most straightforward way of maximizing the detection sensitivity of a faint companion.

\section{Discussion}

Companions to Vega have been searched using several other systems including the Palomar AO system (Metchev et al. 2003), the Keck AO system (Macintosh et al. 2003), and the Gemini AO system (Marois et al. 2006, hereafter M06).  At 7 arcseconds, the distance of a likely perturbing companion for the cold dust, Metchev et al. estimate a 5 sigma limit of H=14.2, which corresponds to a 19 M$_J$ planet using the models of Burrows et al. (1997).  For Macintosh et al. the limiting magnitude at K band is K=17.3 or 10 $M_J$.  M06 made significant improvement on the H band limit of Metchev et al. through the use of Angular Differential Imaging (ADI).  They present a 12,000 s observation which achieves a 5 sigma limit of H=19.5 at 7 arcseconds, an impressive limit which is equivalent to a 4 M$_J$ planet.  

Analysis of the Clio data, as shown in Figures 1 and 2 suggest that at 7 arcseconds separation we are constrained by background noise to a limit of about 0.8 mJy (13.3 magnitudes), or 7 M$_J$ for an approximately 11 minute observation.  This limit is similar for separations as close as approximately 2.5 arcseconds from the star.  The background noise appears random in this region allowing us to extrapolate the expected sensitivity for longer integrations.  For example, a similarly long exposure of 12,000 seconds at M band can reasonably be expected to achieve a limiting magnitude of 0.18 mJy (14.8 magnitudes), corresponding to a companion limit of approximately 2.5 M$_J$ around Vega, and which would be detectable as close as a few arcseconds from the star. 

\subsection{Implications for Planet Detection}

The relative contrast versus separation achievable in M band presented in Figure 4 is similar to what has been achieved with either HST or the best ground-based adaptive optics systems in the near infrared.  For example M06 demonstrate a limit of 11.1-11.9 magnitudes at 0.8 arcsecond separation on Gemini using the Altair system at H band.  Using spectral difference imaging on the VLT a 5 sigma limit of 11.0 has been achieved (Biller et al. 2005). Figure 4 shows a 5 sigma limit of 11 at one arcsecond.  Scaled to the same telescope diameter, these three observational results have similar limiting contrast levels.

In addition to the improved PSF stability the relative contrast of Jupiter-like companions at longer wavelengths provides a significant advantage for the approach outlined here.  Burrows et al. (2003) predict the contrast for a 10 M$_J$ planet around Vega is 15.9 magnitudes at H band, while it is only 12.7 at M band - a difference of 3.2 magnitudes.  For a 3 M$_J$ planet the H and M band magnitudes are predicted to be 20.2 and 14.7 respectively.  For close separations where PSF subtraction dominates the achievable contrast, observing at longer wavelengths can lead to detection of significantly less massive companions.  To be clear, the current data presented here do not yet demonstrate detectability at this level, but the PSF subtractions carried out show that the contrasts are reachable in tandem with observations of an object spanning a significant change in parallactic angle.

The observations presented here illustrate that an optimized imager, when used with an integrated adaptive optics system, can provide detection capabilities similar to, or even better than near-infrared techniques. The useful detection for M band is outside of approximately 3 $\lambda$/D, or 0.5 arcseconds for the MMT where a contrast of approximately 9 magnitudes is achieved.  Inside this region the shorter wavelengths could likely be used to detect relatively bright companions ($\Delta$m$<$5) which would be more difficult to extract at the lower spatial resolution achieved at M band.  Outside of this region, the observations presented here show that 3-5 micron imaging can provide similar contrast ratios versus separation. Coupled with the more favorable contrast in the spectral window, this makes the search for planetary mass companions at longer wavelengths an attractive alternative to near infrared observations.

Although L' and M band provide interesting spectral regions to find planets, optimum candidate stars for observations at NIR versus L' and M will likely be quite different. For H band typical surveys have focused on very young stars ($<$300 Myr) in order to be able to detect the planets while they are still relatively hot and thus bright in the H band.  With an L' and M band survey the most attractive stars become the most nearby stars, although youth is still important.   Several examples of this are detailed in Table 2, for typical targets in H and M band. A 10 M$_J$ planet in a wide orbit around a 5 Gyr old solar twin at 5 pc (M$_s\simeq$H$_s\simeq$2) would have M$_p$=14.2 (Baraffe et al. 2003, Ba03).  The contrast of 12.2 magnitudes would allow detection, from Figure 4, outside of 1.2 arseconds or 6 AU.  The H band flux from Ba03 would be H$_p$=22.4, a contrast of 20 magnitudes which is detectable outside of 8 $\arcsec$ or 40 AU according to the ADI contrast limit of M06.  For a 0.5 Gyr old G star at 20 pc (M$_s\simeq$H$_s\simeq$=5)  Ba03 predicts a 10 M$_J$ planet to be M$_p$=14.9, a contrast of 9.9 magnitudes which would be detectable outside of 0.8 arcseconds, or 16 AU.  H band flux for such a planet is H$_p$=19.2, a contrast of 14.2 magnitudes which is detectable outside of 1.5 arseconds or 30 AU according to the ADI limit of M06. If we consider an even younger solar twin at 50 pc and an age of 0.1 Gyr a 10 M$_J$ planet, at M$_p$=15.5 would not be detectable at M band, and for H band be detectable outside 0.8 $\arcsec$ or 81 AU for H band. These examples show the general trend:  H band detections on large aperture telescopes are best suited to young stars (which are necessarily further away) while 3-5 micron detection is likely to be more sensitive for nearby stars.

The parameter space of detection for M versus H band can be visualized by plotting the limiting distance and age a planet can be detected for various orbits around a solar twin.  Figure 5 shows two plots which illustrate this for a 10 M$_J$ and a 3 M$_J$ planet.  The contrast limits are taken from Figure 4, for M band and from M06 for H band. A sky background limit of M=14.8 and H=23.5 is also assumed.   Planet fluxes in each band are taken from Ba03. M band is preferable to H band on the plots unless the planet is in a wide (~40 AU) orbit or is a young system ($<$ 0.1 Gyr).

Planet detection in the L' and M atmospheric windows requires a large aperture telescope and adaptive optics system as do techniques at shorter wavelengths. However, for L' and M band observations, a noise floor at larger separations arises from the sky background.  For the parameters of our observations of Vega this arises at approximately 10 $\lambda$/D. Thus for L' and M band observation a large aperture benefits not only the sharper image quality (and thus a closer inner working distance) but also fainter limiting sensitivity for companions at larger separations. This makes AO imaging at L' and M band increasingly favorable for larger aperture telescopes.  For example, the Large Binocular Telescope with its 2x8.4 m aperture in coherent imaging mode will have 3.4 times the collecting area and concentrate the light into a PSF which has an angular area 3.7 times less than that of the MMT PSF.  This translates to a limiting magnitude improvement of 2.7. Extrapolating from the MMT sensitivity, we could expect to be sensitive to 2M$_J$ planet around a 1 Gyr old star at 10 pc. 

Next generation telescopes, such as the Thirty Meter Telescope\footnote[1]{see http://www.tmt.org} or the Giant Magellan Telescope\footnote[2]{see http://gmto.org}, can improve these limits further through closer inner working distances and fainter sensitivities, but it will be important for these telescopes to have adaptive optics which are integrated into the telescope, such as the deformable secondary approach being planned for the GMT.

\section{Conclusion}
We present initial M band observations of Vega with a newly commissioned thermal infrared camera, Clio, on the 6.5 MMT using the integrated deformable secondary.  The observation constrains a planet to be less than 7 Jupiter masses in the orientation expected from the dust structure into approximately 2.5 arcsec (20 AU) from the star.    Extrapolation to longer integration times suggests that observations similar to these will be sensitive to wide companion planets at the top end of the mass distribution seen by radial velocity searches, for nearby stars.  

\acknowledgments

Andy Breuninger was instrumental in developing and troubleshooting the Clio electronics and detector.  We thank Vidhya Vaitheeswaren for her dedication in ensuring the deformable secondary AO system operated routinely before and during these observations    Clio is supported by grant NNG04-GN39G from the NASA Terrestrial Planet Finder Foundation Science Program. This work is also supported by the
NASA through the NASA Astrobiology Institute under Cooperative Agreement No.
CAN-02-OSS-02 issued through the Office of Space Science.

\clearpage

\begin{figure}
\epsscale{1.0}
\plottwo{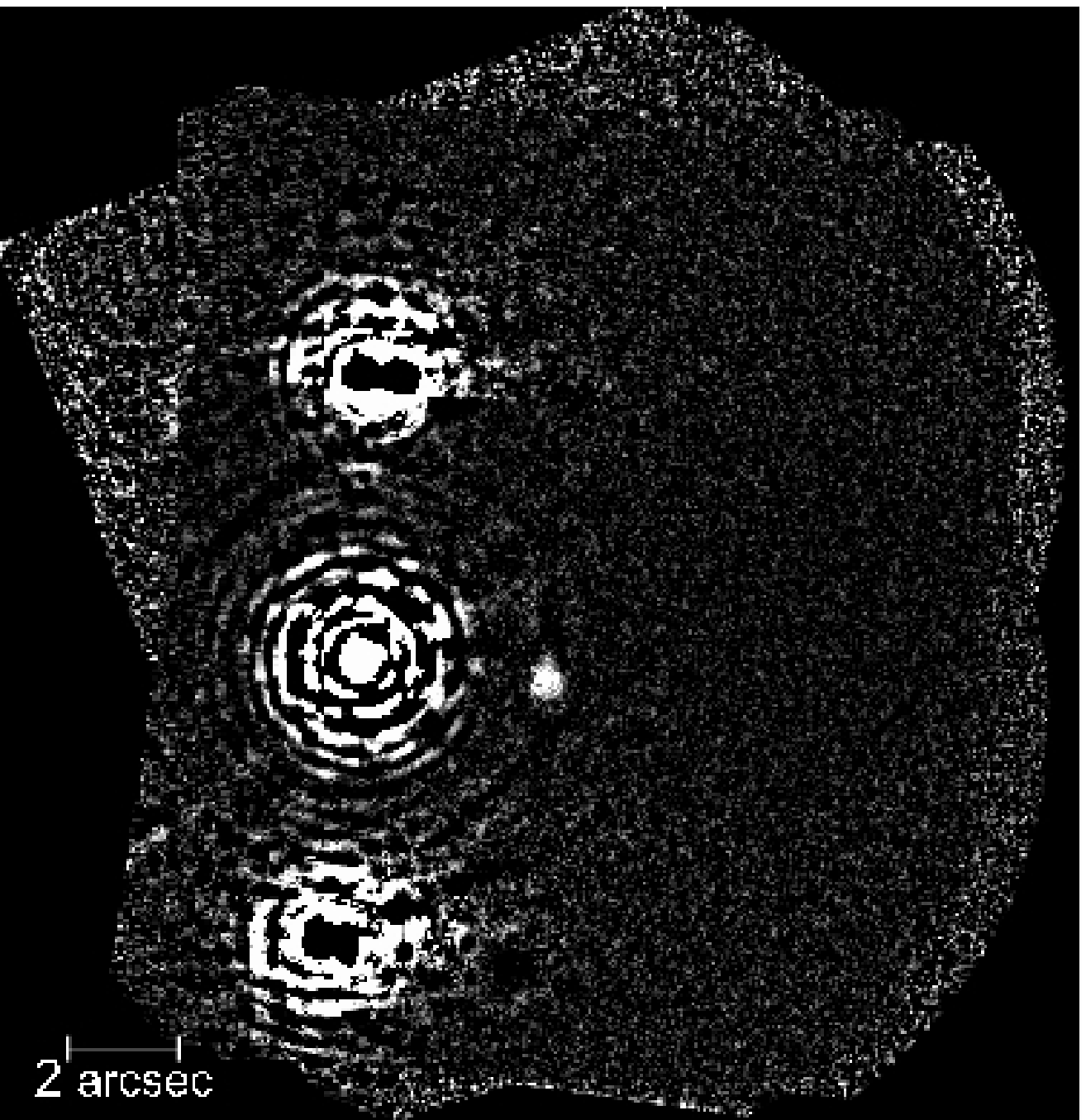}{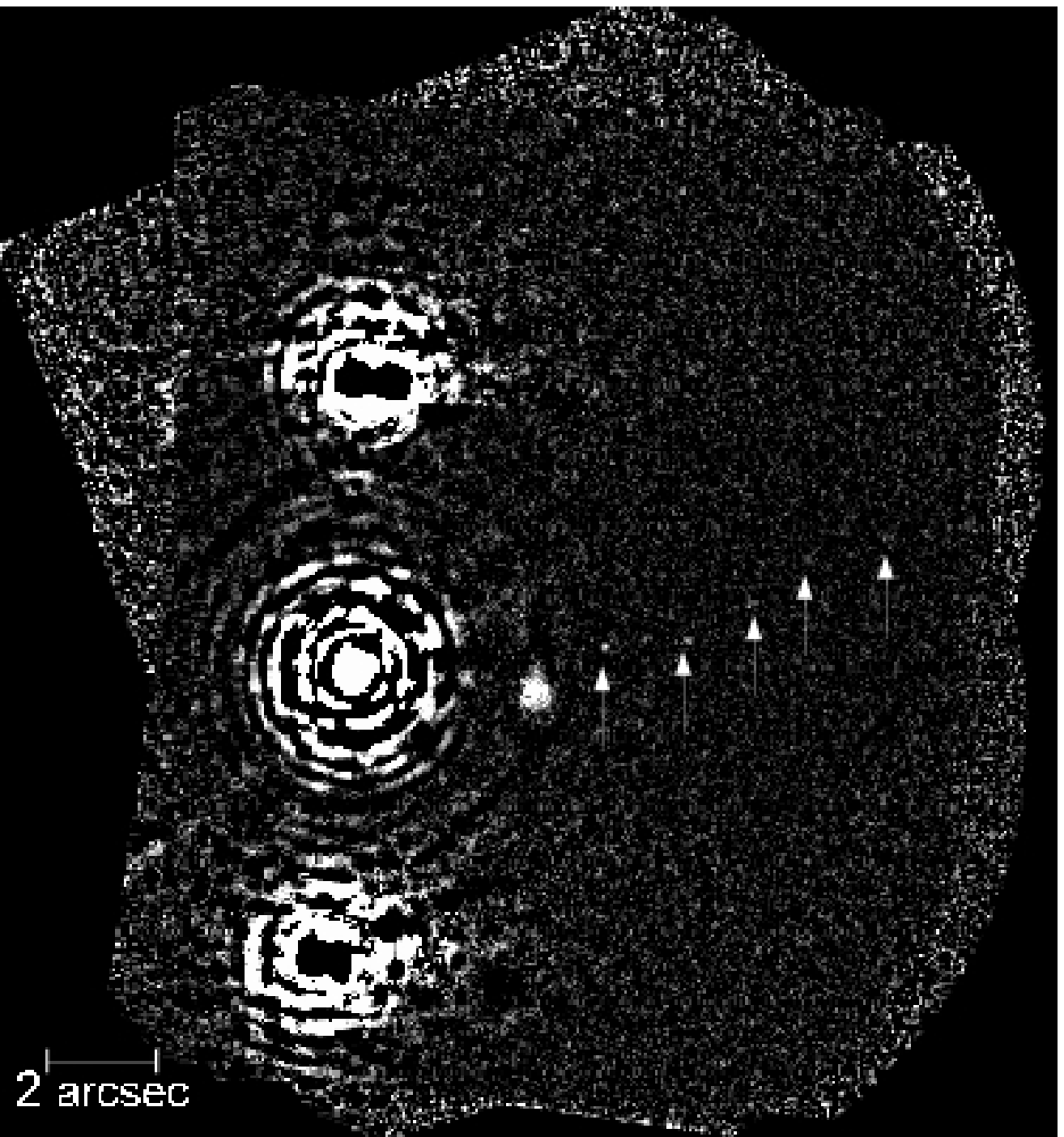}
\caption{Coadded image of Vega.   The image on the left shows a coadded image with a total integration time of 672 seconds.  The image has been unsharp masked, which reveals the Airy pattern diffraction as well as a ghost image to the right of the star.  The artifacts approximately 5 arcseconds above and below the star are residuals of the nod subtraction. North is up and East to the left. No planets are seen to the limit of our sensitivity.  The planet hypothesized by W02 would be in the upper right quadrant of this image.  The right frame is the same with five synthetic planets placed in the raw data so that they appear at the position of the arrows. From left to right the planets are the brightness expected for a 10, 7, 7, 7, and 6 M$_J$ planet.  The rightmost planet is formally at our sensitivity limit, although it cannot be seen reliably in the image. }
\end{figure}

\clearpage

\begin{figure}
\epsscale{.80}
\plotone{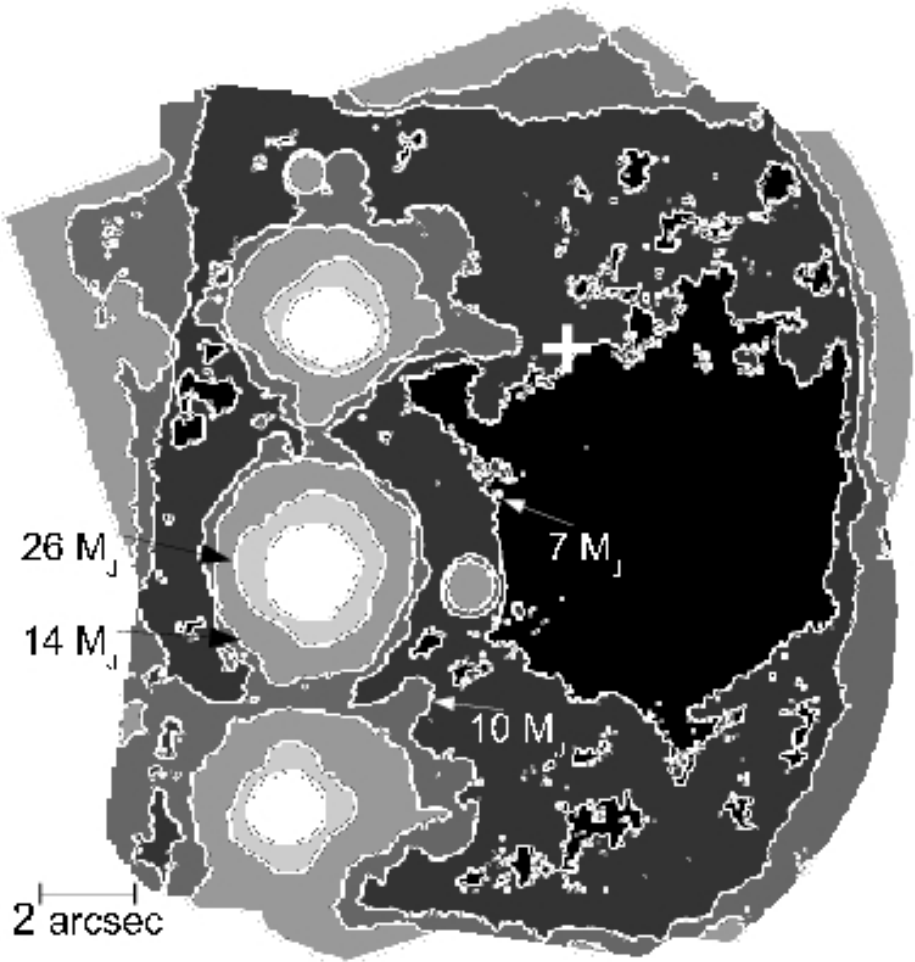}
\caption{Sensitivity image around Vega. The contours show the approximate sensitivity limit for different regions of the image.  From brightest to faintest the contours correspond to sensitivity limits (and corresponding mass limits) of 100 mJy, 16 mJy (26 M$_J$), 2.54 mJy (14 M$_J$), 1.72 mJy(10 M$_J$), and 1.13 mJy (7 M$_J$) .  A cross is drawn at 7 arcseconds northwest of Vega, the predicted position of a planet by W02. }
\end{figure}

\clearpage

\begin{figure}
\includegraphics[angle=-90,scale=.70]{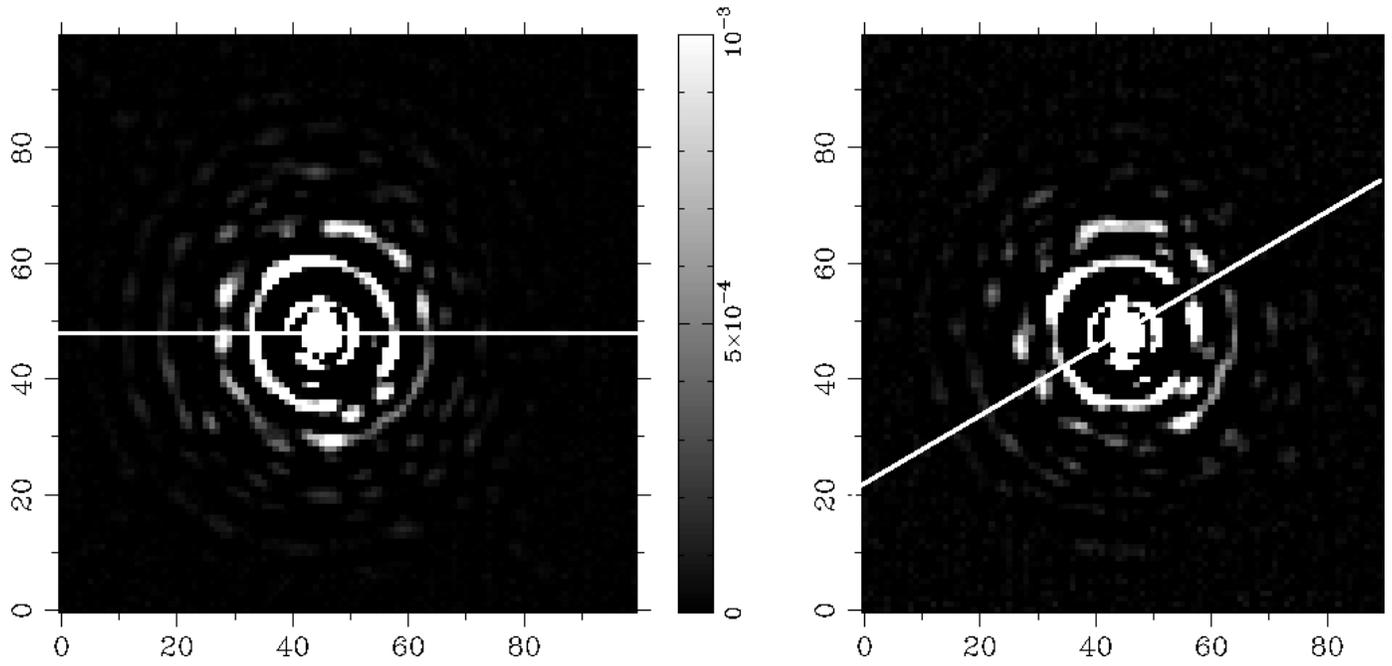}
\caption{Coadded images taken at rotator offsets of 0 and -30 degrees.  Structure in the PSF is very similar between the two, and rotates along with the rotator offset.  This indicates that the aberrations leading to the structure are in the telescope optics.  For this reason "rolling" the telescope by using the derotator will not help remove the PSF structure. }
\end{figure}

\clearpage

\begin{figure}
\epsscale{.80}
\includegraphics[angle=-90,scale=.70]{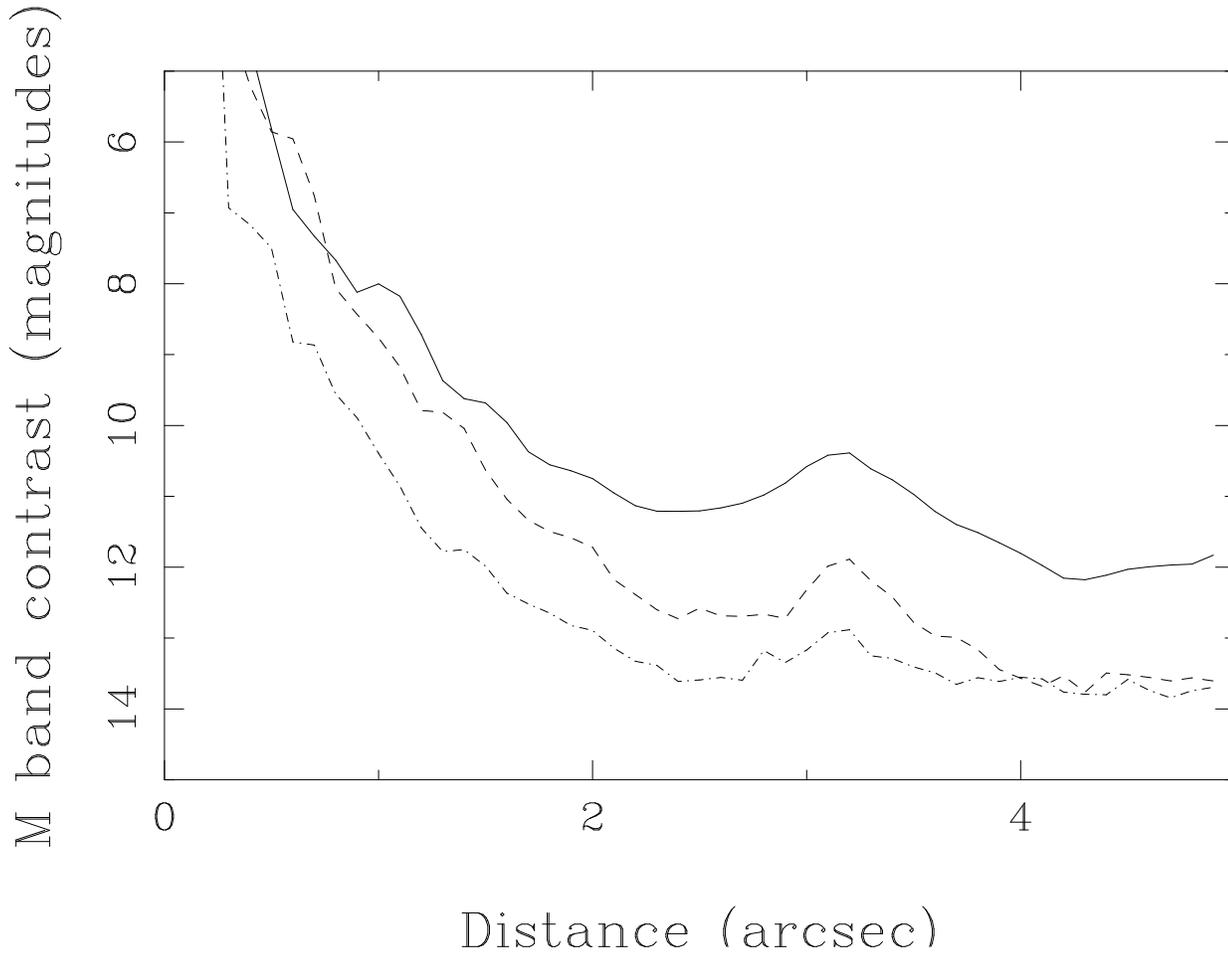}
\caption{Contrast limit versus separation. The 5 sigma contrast limit in M band magnitudes are plotted for Clio on the MMT for a PSF image (solid line), an unsharp masked image (dashed line) and a PSF-subtracted image  (dash-dot line).}
\end{figure}

\clearpage

\begin{figure}
\epsscale{1.0}
\plottwo{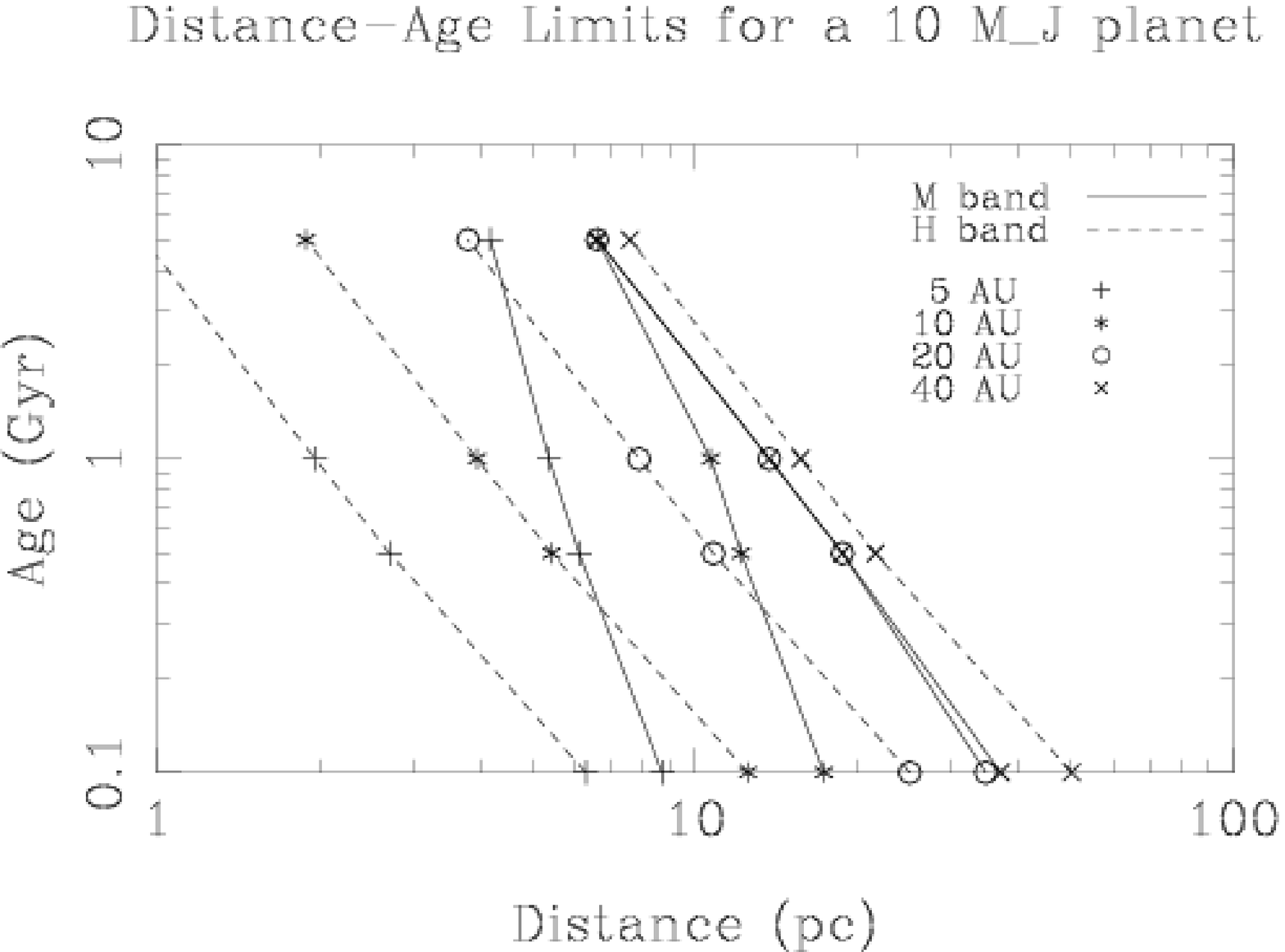}{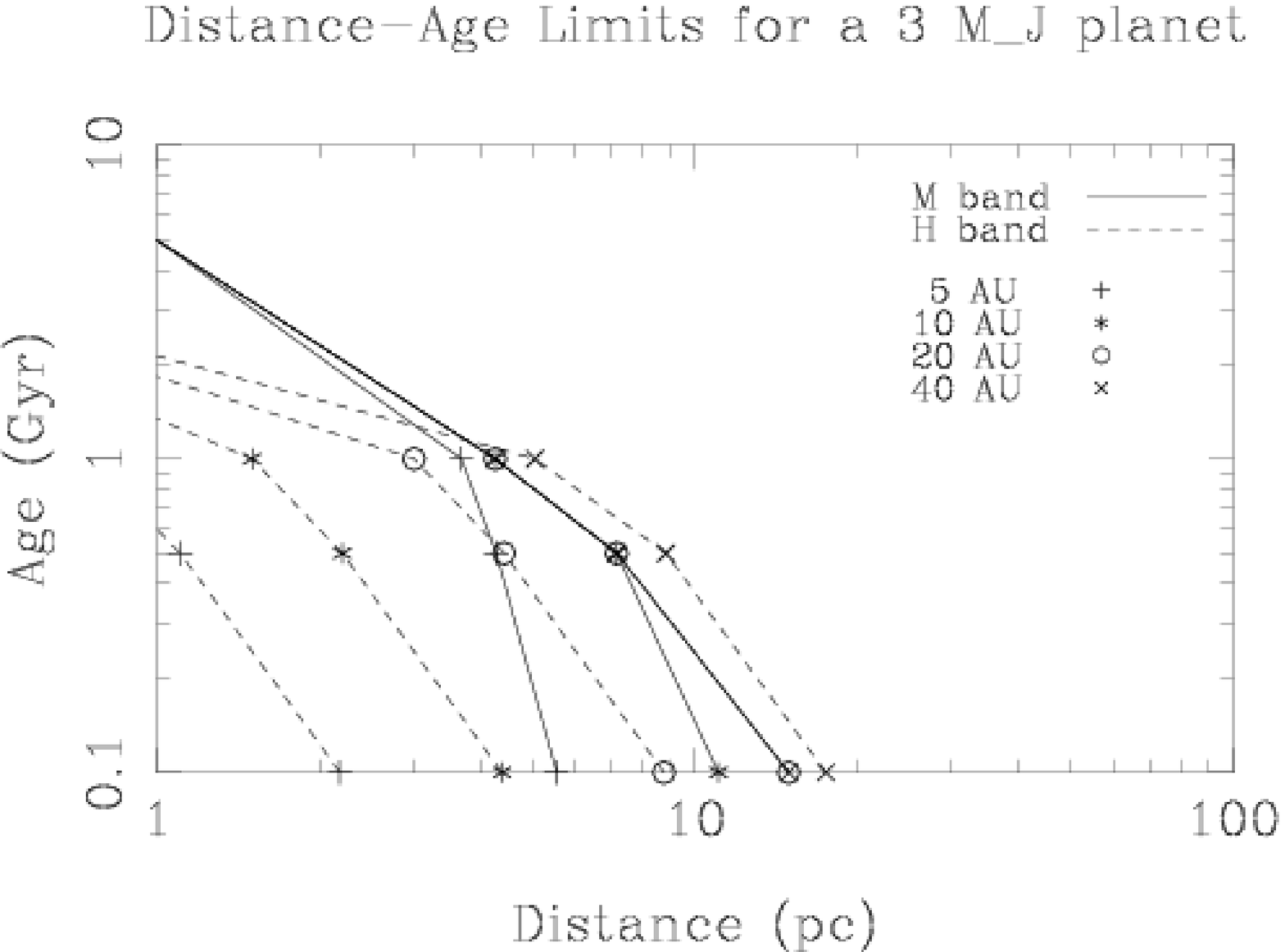}
\caption{Distance-Age limits for 10 and 3 M$_J$ planets around a solar twin.  Limits are shown for orbits of 5, 10, 20, and 40 AU. The contrast versus separation for M band are taken from Figure 4, with the exception that a sky background limit of M=14.8  is used for larger separations.  The contrast versus separation for H band is from M06, with a sky background limit of H=23.5 assumed. }
\end{figure}

\begin{deluxetable}{ccrrrr}
\tabletypesize{\scriptsize}
\tablecaption{Log of data acquisition for Clio observations of Vega June 21, 2005}
\tablewidth{0pt}
\tablehead{
\colhead{Frames} & \colhead{Integration Time (s)}  & \colhead{Position} & \colhead {Parallactic Angle} & \colhead{Rotator Offset} &  \colhead{Comment}
}
\startdata 
1-10	&	12	&	upper left	&-267 	&0 	&	 \\
11-20	&	12	&	lower left	&-268 	&0  	& No AO last two\\
21-22	&	12	&	lower left	&-270 	&0  	&\\
23-24	&	12	&	upper left	&-270 	&0  	&\\
25-26	&	12	&	lower left	&-271 	&0  	&\\
27-28	&	12	&	upper left	&-271 	&0  	&\\
29-30	&	12	&	lower left	&-271 	&0  	&\\
31-32	&	12	&	upper left	&-272 	&0  	&\\
33-34	&	12	&	lower left	&-272  	&0 	&\\
35-36	&	12	&	upper left	&-273  	&0 	&\\
37-38	&	12	&	lower left	&-273  	&0 	&\\
39-40	&	12	&	upper left	&-273 	&0  	&\\
41-42	&	12	&	lower left	&-273 	&0  	&\\
43-44	&	12	&	upper left	&-273 	&0  	&\\
45-46	&	12	&	lower left	&-273 	&0  	&\\
47-48	&	12	&	upper left	&-273 	&0  	&\\
49-50	&	12	&	lower left	&-273  	&0 	&\\
51-52	&	12	&	upper left	&-276 	&-30  	&\\
53-54	&	12	&	lower left	&-276  	&-30 	&\\
55-56	&	12	&	upper left	&-276  	&-30 	&\\
57-58	&	12	&	lower left	&-276  	&-30 	& No AO\\
59-60	&	12	&	lower left	&-276 	&-30  	& No AO\\
61-62	&	12	&	lower left	&-276 	&-30  	&\\
63-64	&	12	&	upper left	&-276 	&-30  	&\\
65-66	&	12	&	lower left	&-276 	&-30  	&\\
67-68	&	12	&	upper left	&-276 	&-30  	&\\
69-70	&	12	&	lower left	&-276  	&-30 	&\\
\enddata

\end{deluxetable}

\clearpage

\begin{deluxetable}{cccccccc}
\tabletypesize{\scriptsize}
\tablecaption{Expected minimum detection angle for a 10 M$_J$ planet  around a solar twin at H and M band}
\tablewidth{0pt}
\tablehead{
\colhead{Band} & \colhead{Distance}  & \colhead{Age} & \colhead {Star Magnitude} & \colhead{Planet Magnitude} &  \colhead{Contrast} & \colhead{Min. Sep.} & \colhead{Min. Sep.} \\ 
\linebreak & (pc) & (Gyr) &   &   &  & (arcsec) & (AU)  }
\startdata
M &	5	&	5	&	2 &	14.2 &	12.2 & 1.2 & 6 \\
H &	5	&	5	&	2 &	22.4 &	20   &  8 & 40 \\
\tableline
M &	20	&	0.5	&	5 &	14.9 &	9.9  & 0.8 & 16 \\
H &	20	&	0.5	&	5 &	19.2 &	14.2 & 1.5 & 30 \\
\tableline
M &	50	&	0.1	&	7 &	15.5 &	8.5  & -- & -- \\
H &	50	&	0.1	&	7 &	18.4 &	11.4 & 0.8 & 81 \\
\enddata

\tablecomments{ The conversion from contrast to minimum separation for M band is from Figure 4.  The conversion for H band is from M06.}
\end{deluxetable}

\clearpage


\begin{thebibliography}{}
\bibitem[Aumann et al.(1984)]{1984ApJ...278L..23A} Aumann, H.~H., et al.\ 
1984, \apjl, 278, L23 
\bibitem[Backman \& Paresce(1993)]{1993prpl.conf.1253B} Backman, D.~E., \& 
Paresce, F.\ 1993, Protostars and Planets III, 1253 
\bibitem[Baraffe et al.(2003)]{2003A&A...402..701B} Baraffe, I., Chabrier, 
G., Barman, T.~S., Allard, F., \& Hauschildt, P.~H.\ 2003, \aap, 402, 701 
\bibitem[Biller et al.(2005)]{2005prpl.conf.8429B} Biller, B., et al.\ 
2005, Protostars and Planets V, 8429 
\bibitem[Boss(2006)]{2006ApJ...637L.137B} Boss, A.~P.\ 2006, \apjl, 637, 
L137 
\bibitem[Brusa et al.(2003)]{2003SPIE.4839..691B} Brusa, G., et al.\ 2003, 
\procspie, 4839, 691 
\bibitem[Burrows et al.(1997)]{1997ApJ...491..856B} Burrows, A., et al.\ 
1997, \apj, 491, 856 
\bibitem[Burrows et al.(2003)]{2003ApJ...596..587B} Burrows, A., Sudarsky, 
D., \& Lunine, J.~I.\ 2003, \apj, 596, 587
\bibitem[Charbonneau et al.(2005)]{2005ApJ...626..523C} Charbonneau, D., et 
al.\ 2005, \apj, 626, 523 
\bibitem[Chauvin et al.(2004)]{2004A&A...425L..29C} Chauvin, G., Lagrange, 
A.-M., Dumas, C., Zuckerman, B., Mouillet, D., Song, I., Beuzit, J.-L., \& 
Lowrance, P.\ 2004, \aap, 425, L29 
\bibitem[Deming et al.(2005)]{2005Natur.434..740D} Deming, D., Seager, S., 
Richardson, L.~J., \& Harrington, J.\ 2005, \nat, 434, 740 
\bibitem[Freed et al.(2004)]{2004SPIE.5492.1561F} Freed, M., Hinz, P.~M., 
Meyer, M.~R., Milton, N.~M., \& Lloyd-Hart, M.\ 2004, \procspie, 5492, 1561 
\bibitem[Gillett et al.(1969)]{1969ApJ...157..925G} Gillett, F.~C., Low, 
F.~J., \& Stein, W.~A.\ 1969, \apj, 157, 925 
\bibitem[Golimowski et al.(2004)]{2004AJ....127.3516G} Golimowski, D.~A., 
et al.\ 2004, \aj, 127, 3516 
\bibitem[Gratton et al.(2004)]{2004SPIE.5492.1010G} Gratton, R., et al.\ 
2004, \procspie, 5492, 1010 
\bibitem[Heinze et al.(2006)]{2006SPIE.} Heinze, A.~N., Hinz, P.~M. \& Sivanandam, S.\ 2006, \procspie, in press
\bibitem[Holland et al.(1998)]{1998Natur.392..788H} Holland, W.~S., et al.\ 
1998, \nat, 392, 788 
\bibitem[Koerner et al.(2001)]{2001ApJ...560L.181K} Koerner, D.~W., 
Sargent, A.~I., \& Ostroff, N.~A.\ 2001, \apjl, 560, L181 
\bibitem[Kuchner \& Holman(2003)]{2003ApJ...588.1110K} Kuchner, M.~J., \& 
Holman, M.~J.\ 2003, \apj, 588, 1110 
\bibitem[Liou \& Zook(1999)]{1999AJ....118..580L} Liou, J.-C., \& Zook, 
H.~A.\ 1999, \aj, 118, 580 
\bibitem[Lloyd-Hart(2000)]{2000PASP..112..264L} Lloyd-Hart, M.\ 2000, 
\pasp, 112, 264 
\bibitem[Lowrance et al.(2005)]{2005AJ....130.1845L} Lowrance, P.~J., et 
al.\ 2005, \aj, 130, 1845
\bibitem[Macintosh et al.(2003)]{2003SPIE.5170..272M} Macintosh, B.~A., et 
al.\ 2003a, \procspie, 5170, 272 
\bibitem[Macintosh et al.(2003)]{2003ApJ...594..538M} Macintosh, B.~A., 
Becklin, E.~E., Kaisler, D., Konopacky, Q., \& Zuckerman, B.\ 2003b, \apj, 
594, 538
\bibitem[Marois et al.(2000)]{2000PASP..112...91M} Marois, C., Doyon, R., 
Racine, R., \& Nadeau, D.\ 2000, \pasp, 112, 91 
\bibitem[Marois et al.(2006)]{2006ApJ...641..556M} Marois, C., 
Lafreni{\`e}re, D., Doyon, R., Macintosh, B., \& Nadeau, D.\ 2006, \apj, 
641, 556 
\bibitem[Metchev et al.(2003)]{2003ApJ...582.1102M} Metchev, S.~A., 
Hillenbrand, L.~A., \& White, R.~J.\ 2003, \apj, 582, 1102 
\bibitem[Metchev \& Hillenbrand(2004)]{2004ApJ...617.1330M} Metchev, S.~A., 
\& Hillenbrand, L.~A.\ 2004, \apj, 617, 1330 
\bibitem[Miller et al.(2004)]{2004SPIE.5490..207M} Miller, D.~L., Brusa, 
G., Kenworthy, M.~A., Hinz, P.~M., \& Fisher, D.~L.\ 2004, \procspie, 5490, 
207 
\bibitem[Mugrauer et al.(2005)]{2005sao..conf..158M} Mugrauer, M., 
Neuh{\"a}user, R., Guenther, E., Brandner, W., Alves, J., \& Ammler, M.\ 
2005, Science with Adaptive Optics, 158 
\bibitem[Neuh{\"a}user et al.(2005)]{2005A&A...435L..13N} Neuh{\"a}user, 
R., Guenther, E.~W., Wuchterl, G., Mugrauer, M., Bedalov, A., \& 
Hauschildt, P.~H.\ 2005, \aap, 435, L13 
\bibitem[Oppenheimer et al.(1998)]{1998ApJ...502..932O} Oppenheimer, B.~R., 
Kulkarni, S.~R., Matthews, K., \& van Kerkwijk, M.~H.\ 1998, \apj, 502, 932
\bibitem[Oppenheimer et al.(2004)]{2004SPIE.5490..433O} Oppenheimer, B.~R., 
et al.\ 2004, \procspie, 5490, 433 
\bibitem[Riccardi et al.(2002)]{2002bcao.conf...55R} Riccardi, A., et al.\ 
2002, Beyond conventional adaptive optics : a conference devoted to the 
development of adaptive optics for extremely large telescopes.~ Proceedings 
of the Topical Meeting held May 7-10, 2001, Venice, Italy.~Edited by 
E.~Vernet, R.~Ragazzoni, S.~Esposito, and N.~Hubin.~Garching, Germany: 
European Southern Observatory, 2002 ESO Conference and Workshop 
Proceedings, Vol.~ 58, ISBN 3923524617, p.~55, 55 
\bibitem[Roques et al.(1994)]{1994Icar..108...37R} Roques, F., Scholl, H., 
Sicardy, B., \& Smith, B.~A.\ 1994, Icarus, 108, 37 
\bibitem[Rowe et al.(2005)]{2005AAS...20711007R} Rowe, J.~F., et al.\ 2005, 
American Astronomical Society Meeting Abstracts, 207,  
\bibitem[Song et al.(2001)]{2001ApJ...546..352S} Song, I., Caillault, 
J.-P., Barrado y Navascu{\'e}s, D., \& Stauffer, J.~R.\ 2001, \apj, 546, 
352 
\bibitem[Sivanandam et al.(2006)]{2006SPIE.} Sivanandam, S., Hinz, P.~M., Heinze, A.N., \& Freed, M.\ 2006, \procspie, in press
\bibitem[Su et al.(2005)]{2005ApJ...628..487S} Su, K.~Y.~L., et al.\ 2005, 
\apj, 628, 487 
\bibitem[Wildi et al.(2003)]{2003SPIE.4839..155W} Wildi, F.~P., Brusa, G., 
Riccardi, A., Lloyd-Hart, M., Martin, H.~M., \& Close, L.~M.\ 2003, 
\procspie, 4839, 155 
\bibitem[Wilner et al.(2002)]{2002ApJ...569L.115W} Wilner, D.~J., Holman, 
M.~J., Kuchner, M.~J., \& Ho, P.~T.~P.\ 2002, \apjl, 569, L115 
\bibitem[Wright et al.(2004)]{2004ApJS..152..261W} Wright, J.~T., Marcy, 
G.~W., Butler, R.~P., \& Vogt, S.~S.\ 2004, \apjs, 152, 261 
\bibitem[Wyatt(2003)]{2003ApJ...598.1321W} Wyatt, M.~C.\ 2003, \apj, 598, 
1321 
 
\end{thebibliography}
\end{document}